# Formation of Negative Hydrogen Ion in Positronium – Hydrogen Collisions

S. Roy and C. Sinha*

Theoretical Physics Department, Indian Association for the Cultivation of Science,

Kolkata - 700032, India.

* E – mail : chand_sin@hotmail.com

**Abstract:**

The importance of the excited states of Positronium ( Ps ) in the formation cross sections ( both differential and total ) of the negative hydrogen ion ( $H^-$ ) are investigated theoretically for the charge transfer reaction, $Ps\ (n = 1, 2) + H \rightarrow e^+ + H^-$ for a wide range of incident energies ( e. g., threshold – 500 eV ). The calculations are performed in the frame work of a qualitative model, the post collisional Coulomb Modified Eikonal Approximation ( CMEA ). A comparative study is also made between the capture from ground and excited states of the Ps. The present CMEA model takes account of higher order effects which is essential for a rearrangement process where the First Born type Approximation ( Coulomb Born for the ionic case ) is not supposed to be adequate. At low incident energies, the excited states of Ps ( 2s, 2p ) are found to play a dominant role in the $H^-$ formation cross sections. Significant deviations are noted between the present CMEA and the Coulomb Born ( CBA ) results even at very high incident energies ( e. g., $E_i$ = 500 eV ), indicating the importance of higher order effects. At high incident energies the present CMEA differential cross section ( DCS ) exhibits a double peak structure which is totally absent in the CBA and could again be attributed to higher order effects.

## 1 Introduction:

The formation of negative hydrogen ion ( $H^-$ ) finds important practical applications in the field of Astrophysics because of the large concentration of this negative ion ( $H^-$ ) in the transition regions of the planetary nebulae [ 1 ]. Further, it was suggested [ 2 ] that the main source of opacity in the atmosphere of the Sun at red and infrared wavelengths was absorption by the $H^-$ ion. Negative ions also play a major role in a number of areas of Physics & Chemistry involving weakly ionized gases and plasmas [ 1 – 4 ]. Particularly the $H^-$ ion finds important biological applications as it can be used as an efficient antioxidant in human body.

The present study addresses the charge transfer reaction in the simplest Ps – atom system e. g., $Ps(1s, 2s, 2p) + H(1s) \to e^+ + H^-$, where the ground ( 1s ) or the excited ( 2s, 2p ) state Ps breaks up to form a $H^-$ ion in the final channel. In view of the fact that the present process becomes exothermic for higher Ps eigen states ( $n \geq 4$, for the Chandrasekhar [ 4 ] wave function of the $H^-$ ion ) , the importance of the Ps excited states in the $H^-$ ion formation can not be overemphasized. It is thus quite worthwhile to make a comparative study of the $H^-$ ion formation cross sections ( both differential and total ) between the ground and the excited states , particularly in respect of the incident energy and the angular momentum quantum number ( $l$ ). Further, the present theoretical estimates, in absence of any other results , could provide some guidelines to the future experiments involving individual excited Ps states , the latter being already feasible for the Ps formation process [ 5 ] . The excited states of the H atom on the other hand, are expected to be less important than those of the Ps in the $H^-$ ion formation. The importance of the Ps excited states was also noted [ 6 ] in the production of low energy antihydrogen ( $\bar{H}$ ) atoms by collision with antiprotons ( $\bar{p} + Ps \to \bar{H} + e^-$ ) where the $\bar{H}$ production was found to increase rapidly with the excitation of Ps. Although to our knowledge, there exists no experimental work in the literature as yet for collisions involving the excited Ps as target , measurements on the production of excited states Ps were reported much earlier [ 7, 8 ] .

To obtain reliable results at very low incident energies, it is essential to take into account the effect of all possible channels and a coupled channel approach is therefore needed. Therefore, the importance of the inclusion of the above rearrangement channel in the scattering process of the PS + H system in the frame work of a coupled channel approach can not be over emphasized . Further, since the negative hydrogen ion does not possess any discrete excited state, this particular charge transfer process has an added advantage both in respect of experimental and theoretical aspects.

However, despite such paramount importance, theoretical study of this process is quite limited [ 9, 10 ] probably because of the complexity lying with the four body problem. As such , in the absence of any experimental data or any sophisticated theoretical calculations , a comparatively simple but a consistent model could provide a reasonable estimate for this important reaction and provide some guidelines for the future experiments as well as for the sophisticated ( e. g. , CC , R Matrix ) theories.

Theoretically, Ps - atom scattering is one of the most difficult problems since both the projectile and the target are composite objects with internal structures. In the present calculation , the four body problem is dealt within the framework of post collisional Coulomb Modified Eikonal Approximation ( CMEA ) where proper care is taken into account of the long range Coulomb attraction between the outgoing positron and the $H^-$ ion in the final channel. For ionic species, this effect should in no way be neglected to obtain reliable results especially at low and intermediate incident energies. It should be pointed out here that this long range coulomb interaction was totally neglected in the coupled two channel calculations ( for ground state only ) of Biswas [ 9 ] . The present CMEA model takes account of the higher order distortion effects in the asymptotic region as well as in the collisional region and as such unlike the plain eikonal , it can be pushed even up to quite low incident energies . In fact, inclusion of higher order effects is essential for a rearrangement process where the First Born type approximation ( Coulomb Born for the ionic case) is not supposed to be adequate. Further, the present Coulomb Born ( CB ) results , extracted from the present CMEA code could also serve the purpose of reliable inputs to the more sophisticated Close Coupling ( CC ) approximation . However, it should be mentioned here, that at extreme low energies ( near threshold ) the present model might not yield quite reliable results and as already mentioned , a more sophisticated calculation ( e.g., CC or R Matrix ) is needed.

Since the $H^-$ ion can exist only in its ground singlet state ( S = 0 ) , the above charge transfer channel occurs in the electronic spin singlet only. The present work gives special emphasis on electron capture from the excited states of Ps ( 2s, 2p ) since few results for the ground state capture were reported earlier [ 10 ] . To the best of our knowledge, this work is the first theoretical attempt for the negative hydrogen ion formation from the excited states of Ps.

**2 Theory:**

The prior and the post forms of the transition amplitude for the above process are given as

$$T_{if}^{prior} = \langle \Psi_f^- | V_i | \psi_i \rangle \tag{1a}$$

$$T_{if}^{post} = \langle \psi_f | V_f | \Psi_i^+ \rangle \tag{1b}$$

where $V_i$ or $V_f$ is the initial or final channel perturbation which is the part of the total interaction not diagonalized in the initial or final state. In equation (1b) or (1a), $\Psi_i^+$ or $\Psi_f^-$ is an exact solution of the four body problem satisfying the outgoing or incoming - wave boundary condition respectively. The present calculations are performed in the framework of post collisional Coulomb Modified Eikonal Approximation ( CMEA ) as well as in the Coulomb Born Approximation ( CBA ) and as such only the prior form of the transition amplitude $T_{if}$ in equation (1a) is adopted. Thus $\Psi_f^-$ in equation (1a) satisfies the equation

$$(H - E)\Psi_f^- = 0 \tag{2}$$

The asymptotic initial channel wave function $\psi_i$ of equation (1a) is chosen as

$$\psi_i = \phi_{Ps}(|\vec{x} - \vec{r}_1|) \, \phi_H(\vec{r}_2) \, e^{(i\vec{k}_i \cdot \vec{R})} \tag{3}$$

where $\vec{R} = (\vec{x} + \vec{r}_1)/2$ ; $\vec{x}$ and $\vec{r}_1$ be the coordinates of the positron and electron forming the Ps atom ( initially ) with respect to the target nucleus, while the coordinate of the target ( H atom ) electron being denoted as $\vec{r}_2$. $\phi_{Ps}$ stands for the ground ( 1s ) or the excited ( 2s, 2p ) states wave function of the Ps atom [ 11 ] while $\phi_H$ represents the ground - state wave function of H atom. The initial channel perturbation $V_i$ in equation (1a) is given by

$$V_i = \frac{Z_t}{x} - \frac{1}{|\vec{x} - \vec{r}_2|} - \frac{Z_t}{r_1} + \frac{1}{|\vec{r}_1 - \vec{r}_2|} = V_{pT} + V_{eT} \tag{4}$$

where $V_{pT}$ refers to the positron - target interaction while $V_{eT}$ refers to the electron - target interaction. The exact final state wave function $\Psi_f^-$ occurring in the prior form of equation (1a) is approximated in the framework of Coulomb Modified Eikonal Approximation ( CMEA ) as

$$\Psi_f^- = \exp\left(-\frac{\pi\alpha_1}{2}\right) \Gamma(1-i\alpha_1) e^{(i\vec{k}_1 \cdot \vec{x})} {}_1F_1[i\alpha_1, 1, -i(k_1 x + \vec{k}_1 \cdot \vec{x})]$$

$$\exp\left[i\eta_1 \int_z^\alpha \left(\frac{1}{x} - \frac{1}{|\vec{x}-\vec{r}_1|}\right) dz'\right] \phi_H^-(\vec{r}_1, \vec{r}_2) \tag{5a}$$

with $\alpha_1 = -\frac{1}{k_1}$, $\eta_1 = \frac{1}{k_1}$. The corresponding final state wave function in the Coulomb Born approximation (CBA) is given as:

$$\Psi_f^- = \exp\left(-\frac{\pi\alpha_1}{2}\right) \Gamma(1-i\alpha_1) e^{(i\vec{k}_1 \cdot \vec{x})} {}_1F_1[i\alpha_1, 1, -i(k_1 x + \vec{k}_1 \cdot \vec{x})] \phi_H^-(\vec{r}_1, \vec{r}_2) \tag{5b}$$

where $\phi_H^-(\vec{r}_1, \vec{r}_2)$ in equations (5a) & (5b) corresponds to the $H^-$ ion ground-state wave function of Chandrasekhar [4] and is given by:

$$\phi_H^-(\vec{r}_1, \vec{r}_2) = (1/4\pi) N \left(e^{-\alpha r_1} e^{-\beta r_2} + e^{-\beta r_1} e^{-\alpha r_2}\right) \tag{6}$$

with N = 0.3948, $\alpha$ = 1.03925, $\beta$ = 0.28309. The ground state energy of the $H^-$ ion for this wave function [4] is E = -0.513 a. u. Although the present work uses the simpler wave function ($H^-$ ion) of Chandrasekhar [4] that was widely used successfully [9, 12 - 15], more sophisticated wave functions are available in the literature [12, 16] involving explicit correlation (e – e) terms and producing more accurate binding energy of the $H^-$ ion. However, the correlated wave function is expected to be more suitable and in fact unavoidable for the two electron transition processes that are mainly governed by the e – e correlation [17, 18]. Since the present work refers to single electron capture and the $H^-$ ion is formed in the final state, to our belief, the use of this simpler representation of the $H^-$ ion [4] which partly takes account of the e – e correlation [13, 14] is quite justified and is supposed to give reasonable description of the physics of the target for this particular process where the electron is transferred from the Ps atom to the one electron target to form the $H^-$ ion in the final state. In fact, the choice of this simpler wave function is mainly dictated by the feasibility of the present calculation which is already quite involved for a four body problem. Even in the simple first Born approximation ( FBA ) , the mathematical expression for the scattering amplitude becomes quite complex with the correlated wave function [12].

Using equations ( 2 ) - ( 6 ) in equation 1(a) and then after much analytical reduction [ 19 ] , the transition amplitude is finally reduced to a two - dimensional integral to be evaluated numerically [ 19, 20 ]. The differential cross section for the process studied is given by

$$\frac{d\sigma}{d\omega} = \frac{1}{4} \frac{v_f}{v_i} \left| T_{if} \right|^2 \qquad (7)$$

where $v_i$ and $v_f$ are the initial and the final velocities of the Ps and the $e^+$ respectively. The factor of $\frac{1}{4}$ in eqn.( 7 ) is introduced to account for the singlet ( para ) state formation of $H^-$. The ortho state of $H^-$ being much unstable, does not contribute to the $H^-$ formation cross sections [ 9 ]. The total cross section $\sigma$ is obtained by integrating eqn. ( 7 ) over the solid angle of the outgoing positron, i. e.,

$$\sigma = 2\pi \int_0^\pi \frac{d\sigma}{d\omega} \sin\theta \, d\theta \qquad (8)$$

**3 Results & Discussions:**

We have computed the results for the $H^-$ ion formation cross sections , both differential ( DCS ) and total ( TCS ) , for the charge transfer reaction Ps ( 1s, 2s, 2p ) + H ( 1s ) → $e^+$ + $H^-$ ( 1s ) in the Coulomb Modified Eikonal Approximation ( CMEA ) . For comparison , the corresponding Coulomb Born approximation ( CBA ) results are also extracted from the present CMEA code .

Since the threshold energy for this particular process is 6.45 eV for the ground ( n = 1 ) Ps state and 1.35 eV for the excited ( n = 2 ) Ps states, the DCS results at a much lower incident Ps energy ( e.g., 2 eV or 5 eV ) are exhibited in figure 1 for the excited states ( 2s, 2p ) of Ps only. Figures 1 – 3 exhibit the present differential cross sections ( DCS ) from the ground and excited ( 2s, 2p ) states of Ps for incident energies ( $E_i$ ) 2 eV, 5 eV, 10 eV and 30 eV respectively.

Figures 1 – 3 reveal that the $H^-$ ion formation is strongly favoured in the forward direction for all the states of Ps , although the maximum DCS ( at $\theta = 0^\circ$ ) that mainly governs the magnitude of the total formation cross section ( TCS ) , is largest for the 2p state and smallest for the 2s one , while the 1s state lies in between , i.e., $2p \succ 1s \succ 2s$ . In contrast , for higher scattering angles , the partial DCS follows the order 1s > 2p > 2s ( vide fig. 2 ) . This trend of the DCS at forward angles is noted even up to 25 eV ( not shown in figure ) , although with increasing

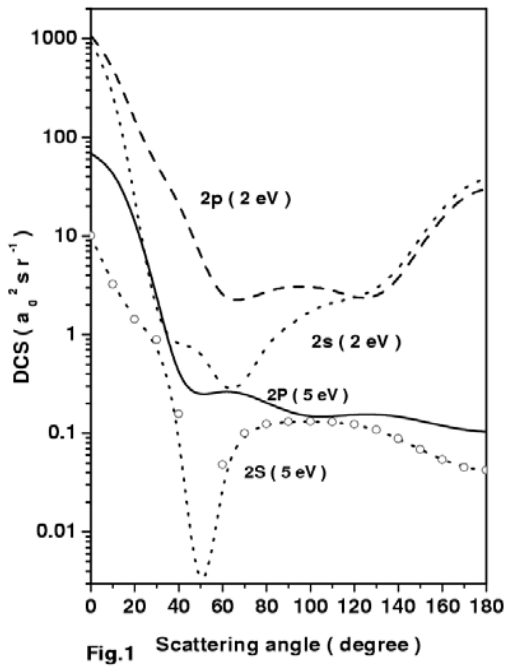

Fig.1 . Differential cross sections ( DCS ) ( in $a_0^2\ sr^{-1}$ ) for the $H^-$ ion formation in terms of the scattering angle $\theta$ of the outgoing positron in Ps + H atom collisions for incident energy ( $E_i$ ) = 2 eV and 5 eV. Dashed line for 2p state , dotted line for 2s state of the Ps atom for ( $E_i$ ) = 2 eV. For $E_i$ = 5 eV, solid line for 2p state , dotted line with open circle for 2s state of Ps atom.

incident energy, the maximum of the 2p DCS decreases and tends towards the 1s maximum ( vide fig. 3 ) . As for the qualitative comparison , the 1s , 2s DCS , ( figs 2 & 3 ) exhibit a distinct minimum followed by a broad maximum ( or a shoulder like structure ) depending on the incident energy. Both the minimum and the maximum sharpen and move towards the forward direction with increasing $E_i$ . For the 2p state on the other hand , instead of a distinct minimum , only a kink followed by another broad minimum are noted ( to be explained later ) . Here also the positions and the sharpness of the structures are dependent on the incident energy .

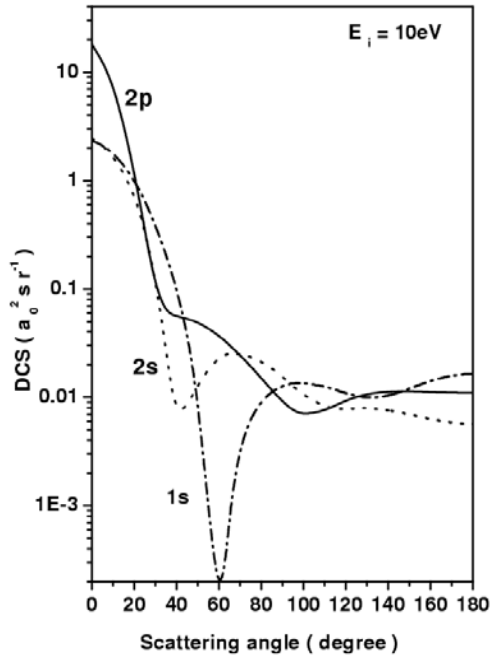 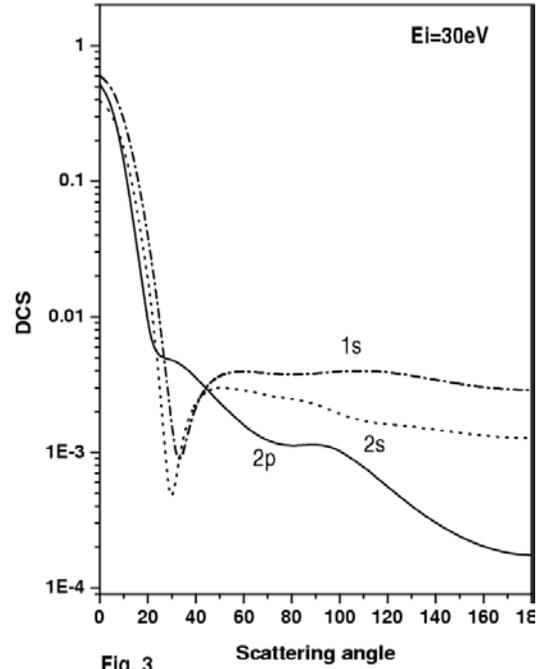

Fig. 2. Differential cross sections ( DCS ) ( in $a_0^2 \, sr^{-1}$ ) for the $H^-$ ion formation in terms of the scattering angle θ of the out going positron in Ps + H atom collisions from the 1s, 2s, 2p states of Ps atom for incident energy ($E_i$) = 10 eV, solid line for 2p state, dotted line for 2s state, dash dot line for 1s state of the Ps atom.

Fig. 3. Same DCS as in fig. 2 but for $E_i$ = 30 eV.

Figs. 4 ( a ) - 4 ( c ) exhibit the CMEA DCS ( 1s, 2s & 2p ) at Ps energy 15 eV along with the individual contributions to the total DCS due to the net attractive and repulsive parts of the total prior interaction, $V_i$ in equation ( 4 ). The net attractive interaction that governs mainly the soft collision, strongly dominates ( by orders of magnitude ) over the repulsive one at forward angles and as such determines the nature of the full DCS.

The occurrence of the distinct minimum in the 1s, 2s DCS ( vide figs 1 - 4 ( b ) ) may be ascribed to the destructive interference between the attractive and the repulsive parts of the scattering amplitude at this particular angle . Since with increasing $E_i$, the amplitude for the attractive part becomes more and more peaked in the forward direction , the position of the DCS minimum shifts towards smaller angles with increase of impact energy .

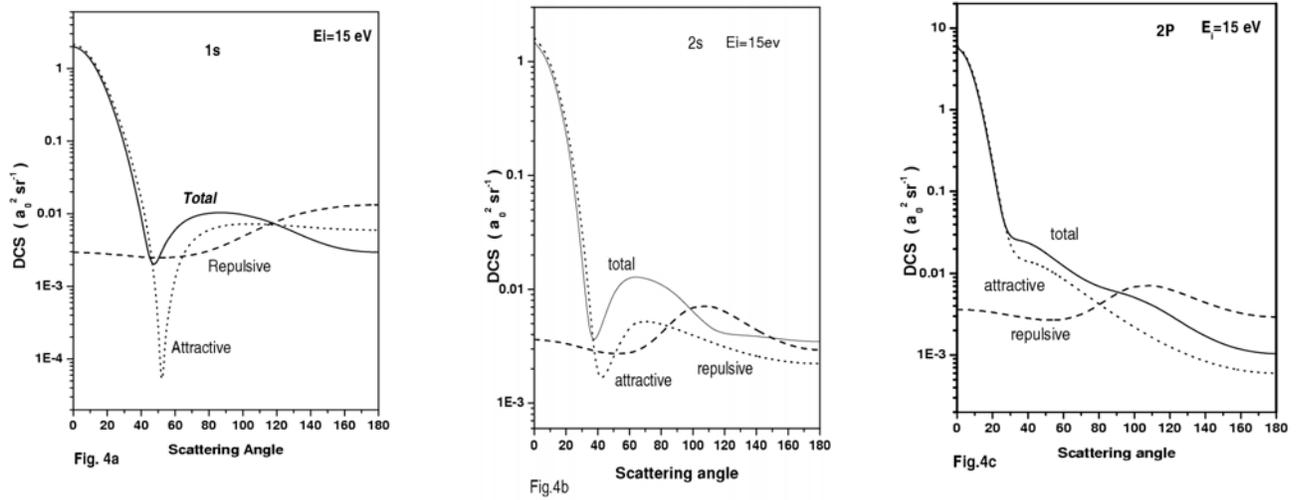

Fig. 4. ( a ) Differential cross sections ( DCS ) ( in $a_0^2\ sr^{-1}$ ) for the $H^-$ ion formation in terms of the scattering angle θ of the out going positron in Ps + H atom collisions for $E_i$ = 15 eV, from 1s state of Ps , solid curve for full interaction potential $V_i$ in equation ( 6 ) , dotted curve for the DCS calculated with the net attractive part [ $V_{eT}$ in eqn. ( 6 )] of the interaction potential $V_i$ at $E_i$ = 15 eV, dashed curve for the DCS calculated with the net repulsive part [ $V_{pT}$ in eqn. ( 6 ) ] of the interaction potential $V_i$ at $E_i$ = 15 eV. ( b ) same as ( a ) but for 2s state of Ps , ( c ) same as ( a ) but for 2p state of Ps atom.

The absence of the deep minimum in the 2P DCS ( fig. 4c ) is attributed to the fact that the minima for the m degenerate state occur at different scattering angles , i. e., the maxima and minima of the $2p_0$ ( m = 0 ) state occur at different scattering angles than those for the $2p_{\pm 1}$ ( $m = \pm 1$ ) states.

Figures 5 & 6 that display the DCS ( 1s, 2s, 2p ) at some higher incident energies, ( e.g., $E_i$ = 50 , 100, 150 & 200 eV ) reveal that with increasing $E_i$ , a double peak structure starts appearing and as in the case of the first minimum ( vide figs. 1 – 4 ), the position of the secondary peak shifts towards smaller angles. Further, the secondary structures for all the states, become more and more prominent with increasing $E_i$ , indicating the importance of the higher – order effects at higher incident energies . The inset of fig. 5 ( for the 1s state ) gives a more clear evidence of this.

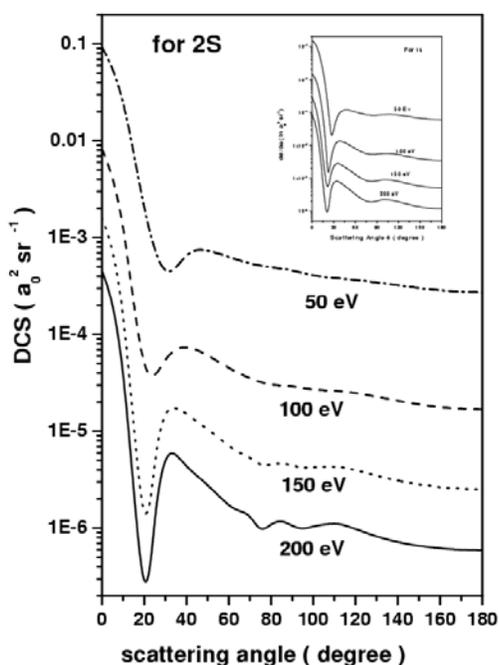

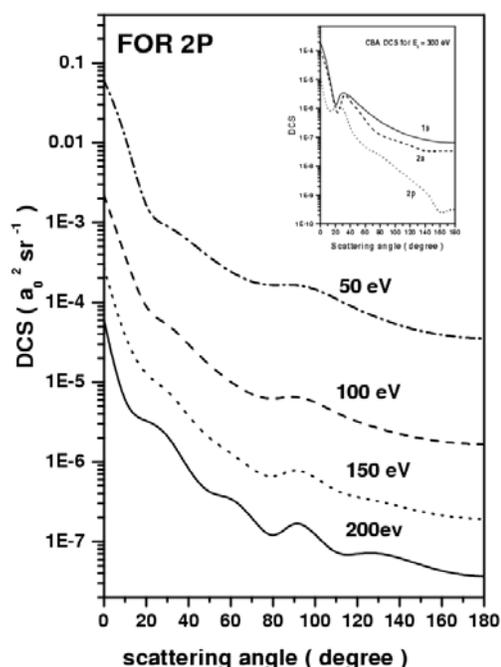

Fig. 5. $H^-$ ion formation cross section ( DCS ) from the 2s state of Ps atom for different incident energies, dashed dot curve for $E_i$ = 50 eV, dashed curve for $E_i$ = 100 eV, dotted curve for $E_i$ = 150 eV, solid curve for $E_i$ = 200 eV. Inset same DCS but from 1s state of Ps.

Fig. 6. Same as Fig. 5 but for 2p state of Ps. Inset CB DCS for $E_I$ = 300 eV. Solid curve for the 1s state of Ps, dashed curve for 2s state of Ps and dotted curve for 2p state of Ps.

The appearance of the double peak structure in the 1s & 2s DCS at higher $E_i$ could be attributed to the higher order effect that is taken into account through the post collisional eikonal approximation. For the 2p state on the other hand, this prominent secondary peak is again somewhat suppressed resulting in some hump like structures ( fig. 6 ) , due to the same reason as stated before for the first minimum . It should be pointed out here that no such secondary structure occurs in the CBA results ( vide the inset of fig. 6 ) even at very high incident energy ( e. g., 300 eV ).

Figures 7 & 8 exhibit the partial ( 1s, 2s, 2p ) total cross sections ( TCS ) for the $H^-$ ion formation for a wide range of incident energies . As is noted from the figures, at very low incident energies ( up to 15 eV ) , the partial TCS follows the order 2p > 1s > 2s while at intermediate energies ( 15 – 25 eV ) , the order is 1s > 2p > 2s . Beyond 25 eV onwards ( up to 200 eV ) the partial TCS is in the decreasing order i.e., 1s > 2s > 2p . The dominance of the 2p TCS at low incident energies could probably be attributed to the long range polarization effects which is much stronger for the 2p state than for any other states. In fact, a major contribution to the polarization effect that mainly dominates at lower incident energies , comes from the lowest lying p state ( i. e., 2p state ).

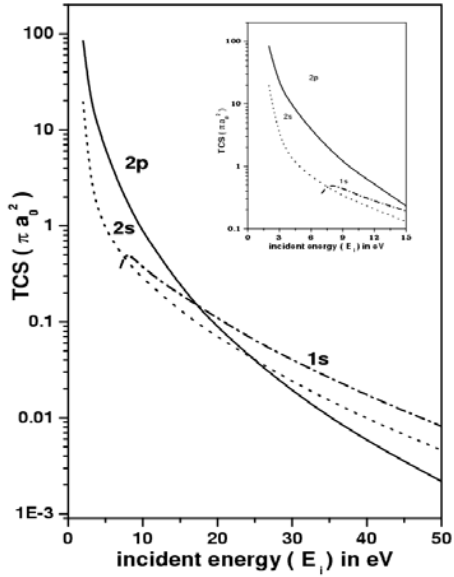

Fig. 7 . Partial total cross sections ( TCS ) ( units of $\pi a_0^2$ ) for $H^-$ formation from 1s, 2s and 2p states of Ps for incident energy range from threshold – 50 eV. Solid curve for 2p state, dotted curve for 2s state, dash dot curve for 1s state. Inset same TCS but for incident energy range threshold – 15 eV.

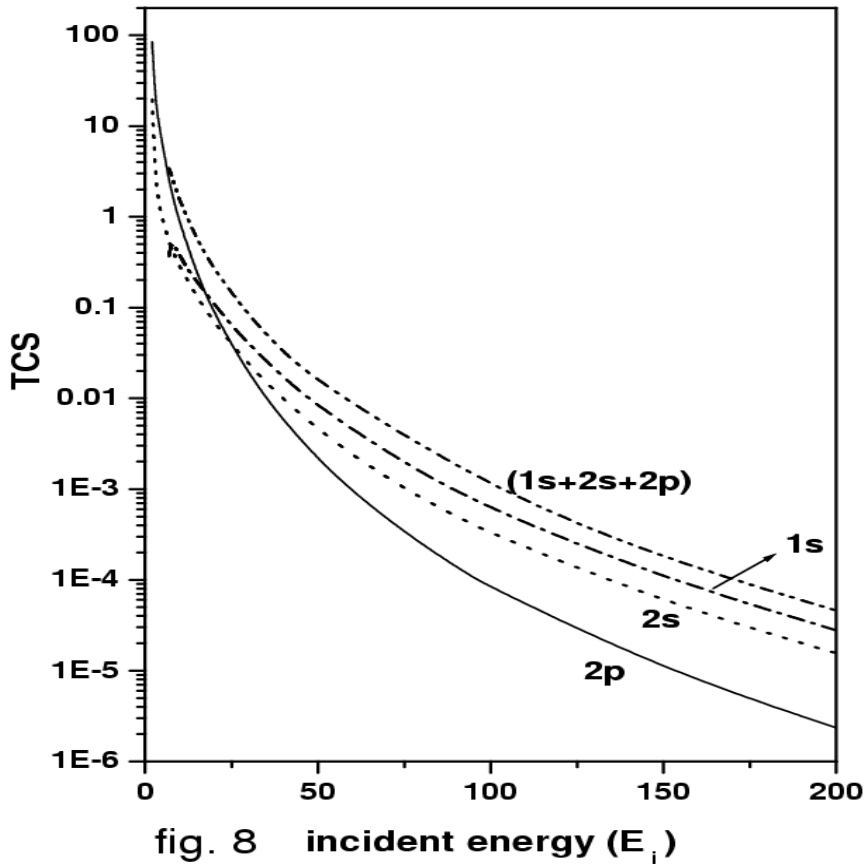

Fig. 8. Partial total cross sections ( TCS ) for $H^-$ formation from 1s, 2s and 2p states of Ps atom and the integrated cross section ( 1s+ 2s + 2p ) for a wide range of incident energy ( up to 200 eV). Solid curve for 2p state, dotted curve for 2s state, dash dot curve for 1s state and dash- double dot curve for integrated results.

Figure 8 also demonstrates the integrated CMEA total cross sections ( 1s + 2s + 2p ) for the incident energies up to 200 eV. As is evident from fig. 8, all the partial TCS and obviously the integrated one decrease monotonically with increasing incident energy.

Table 1: Partial CMEA and Integrated TCS for the $H^-$ ion formation from the ground ( n = 1 ) and the excited ( n = 2 ) states of Ps ( * CBA is derived from the present CMEA code ).

| $E_i$ (eV) | Partial Total Cross Sections ||||||| Integrated Total Cross Sections ||
| | 1s ||| 2s || 2p || 1s + 2s + 2p ||
| | CBA* Present | CMEA Present | CC of Biswas [9] | CBA* Present | CMEA Present | CBA* Present | CMEA Present | CBA* Present | CMEA Present |
|---|---|---|---|---|---|---|---|---|---|
| 2. | - | - | - | 16.55 | 19.51 | 83.09 | 84.67 | 99.64 (2s+2p) | 104.18(2s+2p) |
| 5. | - | - | - | 1.86 | 0.95 | 6.47 | 6.31 | 8.33 (2s+2p) | 7.26 (2s+2p) |
| 10. | 0.77 | 0.37 | - | 0.47 | 0.28 | 1.05 | 0.87 | 2.29 | 1.52 |
| 15. | 0.32 | 0.19 | 0.056 | 0.18 | 0.13 | 0.29 | 0.23 | 0.79 | 0.55 |
| 20. | 0.16 | 0.11 | 0.105 | 0.09 | 0.07 | 0.11 | 0.08 | 0.36 | 0.26 |
| 25. | 0.09 | 0.06 | 0.121 | 0.048 | 0.038 | 0.049 | 0.037 | 0.187 | 0.135 |
| 30. | 0.05 | 0.04 | 0.1137 | 0.03 | 0.02 | 0.024 | 0.018 | 0.104 | 0.078 |
| 40. | 0.02 | 0.017 | 0.075 | 0.012 | 0.01 | 0.008 | 0.006 | 0.04 | 0.033 |
| 50. | 0.01 | 0.008 | 0.04 | 5.6E-3 | 4.60E-3 | 3.17E-3 | 2.18E-3 | 0.02 | 0.01 |
| 60. | 5.5E-3 | 4.3E-3 | 0.02 | 2.9E-3 | 2.4E-3 | 1.4E-3 | 9.6E-4 | 9.8E-3 | 7.6E-3 |
| 70. | 3.1E-3 | 2.4E-3 | 0.014 | 1.7E-3 | 1.3E-3 | 7.2E-4 | 4.7E-4 | 5.5E-3 | 4.17E-3 |
| 80. | 1.9E-3 | 1.4E-3 | 7.1E-3 | 1.0E-3 | 8.0E-4 | 3.9E-4 | 2.5E-4 | 3.3E-3 | 2.45E-3 |
| 100. | 7.9E-4 | 5.9E-4 | 2.4E-3 | 4.2E-4 | 3.25E-4 | 1.34E-4 | 8.30E-5 | 1.34E-3 | 9.98E-4 |
| 150. | 1.45E-4 | 1.04E-4 | - | 7.72E-5 | 5.67E-5 | 1.75E-5 | 1.02E-5 | 2.39E-4 | 1.71E-4 |
| 200. | 4.05E-5 | 2.79E-5 | - | 2.14E-5 | 1.56E-5 | 3.99E-6 | 2.34E-6 | 6.59E-5 | 4.58E-5 |
| 350. | 2.81E-6 | 1.81E-6 | - | 1.64E-6 | 1.15E-6 | 5.26E-7 | 4.46E-7 | 4.97E-6 | 3.41E-6 |
| 500. | 4.59E-7 | 2.84E-7 | - | 3.50E-7 | 2.64E-7 | 2.73E-7 | 2.63E-7 | 10.82E-7 | 8.11E-7 |

For the sake of some numerical measure, we have displayed in table 1 a few partial and integrated ( 1s + 2s + 2p ) CMEA TCS along with the corresponding CB results. The CB results are always found to be higher than the CMEA ones except at the threshold region. Further, the significant discrepancy between the present CMEA ( both in the partial and the integrated TCS ) and the CB results even at a very high incident energy ( e. g. , $E_i$ = 500 eV ) indicates the importance of higher order effects for a rearrangement process. Table 1 also reveals that at lower incident energies, the discrepancy between the CMEA and the CB results is much higher for the 1s and 2s states than for the 2p one, while at higher $E_i$, the reverse is true, i. e. , the deviation is most for the 2p state.

To our knowledge, the only theoretical results available for this particular charge transfer reaction ( for the ground state Ps only ) is due to Biswas [ 9 ] in the framework of coupled two channel approximation using their [ 9 ] model exchange method as well as the abinitio exchange method. As stated before , the large deviations ( vide Table 1 ) noted between the two theories ( present and that of Biswas [ 9 ] ) could probably be attributed to the Author's [ 9 ] neglect of the long range Coulomb attraction between the $e^+$ and the $H^-$ in the final channel. Further, the neglect of this interaction leads to an unphysical behaviour , e. g. , the TCS of Biswas [ 9 ] converge to the FBA results while , the present CMEA TCS tend to converge to the Coulomb Born results at very high incident energies, as is expected physically for the present process leading to an ion formation in the final channel . It should be emphasized that this long range Coulomb interaction between the charged particles should in no way be neglected to obtain reliable results.

## 4 Conclusions :

At low and intermediate incident energies, the excited states of the Ps play a dominant role in the $H^-$ ion formation cross sections while at higher energies, the capture from the ground state Ps dominates. Particularly the Ps 2p state is found to be the dominant process among the three states ( 1s, 2s, 2p ) at very low incident energies.

The distinct double peak structures occurring at higher incident energies ( $E_i$ ) in the 1s, 2s DCS ( which are absent in the CBA ) could be the manifestation of higher order effects. The signature of the double peak becomes more and more prominent with increasing $E_i$ indicating the increasing importance of the higher order effects with incident energy. For the 2p state however, the double peak structure is not so prominent as in the case of 1s, 2s because of the m – degenerate states.

The discrepancy between the present CMEA and the CBA TCS arising due to higher order effects retains even at a very high incident energy as is expected for a rearrangement process.

For a more reliable results at very low incident energies ( near threshold ), the polarization of the Ps atom should be taken into account explicitly and a more sophisticated calculation with proper inclusion of the final channel long range coulomb interaction is highly needed.